\documentclass[12pt]{article}
\usepackage{amsmath,epsfig}
\usepackage{amsmath}
\usepackage{amssymb}
\usepackage{mathrsfs}
\usepackage{amsmath,epsfig}
\usepackage{graphicx}
\usepackage{amsmath,epsfig}
\usepackage{epsfig}
\newsavebox{\PSLASH}
\sbox{\PSLASH}{$p$\hspace{-1.8mm}/}
\newcommand{\PS}{\usebox{\PSLASH}}

\begin{document}

\vspace{4cm}
\begin{center}
{\Large\bf{Study of the Top Quark FCNC}}\\
\vspace{1cm} {\bf M. Mohammadi Najafabadi$\
^{\dagger,}$\footnote{\normalsize{Corresponding author email
address: mojtaba@ipm.ir}}} and {\bf
N. Tazik$\ ^{\ddagger}$} \\
\vspace{0.5cm}
{\sl ${\ ^{\dagger}}$  School of Particles and Accelerators, \\
Institute for Research in Fundamental Sciences (IPM) \\
P.O. Box 19395-5531, Tehran, Iran}\\
and\\
{\sl ${\ ^{\ddagger}}$ Physics Department, Semnan University, Semnan, Iran}\\

\vspace{3cm}
 \textbf{Abstract}\\
 \end{center}

We study the one-loop contribution of the effective flavor
changing neutral couplings (FCNC) $tcZ$ to the charm quark
electric dipole moment. Using the known limits on the top and
charm quarks electric dipole moments, we place limits on
these FCNC anomalous couplings.

\newpage

\section{Introduction}

The standard model (SM) is in a very good agreement with present
experimental data. Nonetheless, it is believed to leave many
questions unanswered, and this belief has resulted in numerous
theoretical and experimental attempts to discover a more
fundamental underlying theory. Various types of experiments may
expose the existence of physics beyond the SM, including the
search for direct production of exotic particles at high energy
colliders. A complementary approach in hunting for new physics is
to examine its indirect effects in higher order processes. Since
top quark is far more massive than other SM fermions, its
interactions may be quite sensitive to new physics originating at
higher scale \cite{beneke}. If there are any deviations from the
SM expectations in the properties of the top quark, they may
indirectly lead to modifications in the anticipated branching
fractions.

In the SM, due to the Glashow-Iliopoulos-Maiani (GIM) mechanism,
the top quark Flavor Changing Neutral Current (FCNC) interactions
are absent at tree level and are extremely suppressed at loop
level. Therefore, the observation of any FCNC top quark process a
smoking gun for new physics beyond the SM.

Within the SM framework, FCNC interactions are induced by the $W$
boson charged current Cabbibo-Kobayashi-Maskawa (CKM) transitions
involving down-type quarks in the loops which are much lighter
than the top quark. In models beyond SM such as minimal
supersymmetric standard model (MSSM) or Technicolor theory,
although the top quark FCNC interactions are also induced at loop
level, they can be greatly enhanced relative to the SM
predictions. For example in MSSM, in addition to the $W$ boson
loops, there are four kinds of loops contributing to the top
quark  FCNC interactions. In MSSM, charged Higgs, chargino,
gluino and neutralino loops contribute to the top quark FCNC
interactions. Theoretical branching ratios of FCNC top quark
decays in various models are presented in Table \ref{Limits}
\cite{MinYang},\cite{aguilar}. It is worth mentioning that at the
LHC, the branching fraction for top FCNC decay, $BR(t\rightarrow
Zq)$, can be measured with the precision of $6.1\times 10^{-5}$
and $3.1\times 10^{-4}$ with the integrated luminosity of 100
fb$^{-1}$ and 10 fb$^{-1}$, respectively
\cite{atlas},\cite{cms},\cite{bern}. There are several studies on
the top quark FCNC which some of them can be found in:
\cite{f1},\cite{f2},\cite{f3},\cite{f4},\cite{f5},\cite{f6},
\cite{f7}.

In this article, our aim is to constraint the top quark FCNC anomalous
couplings, in the process of $t\rightarrow Zc$, using effects induced
by the electric dipole moment (EDM) of top quark on the one loop induced EDM
of the charm quark. In the analysis, we will use the estimated bounds on the
EDM's of top and charm quark to constraint the anomalous couplings.

\begin{table}
\begin{center}
\begin{tabular}{|c|c|c|c|}\hline
  Model      &  SM        &  MSSM   &  Technicolor    \\ \hline
 $BR(t\rightarrow cZ)$ &  $\sim 10^{-14}$  &  $\sim 10^{-6}$  & $\sim10^{-4}$ \\ \hline
   \end{tabular}\label{Limits}
\end{center}\caption{Theoretical branching ratios of FCNC top quark
decays in various models.}
\end{table}

\section{Effective FCNC Lagrangian}
One tool that is often used to describe the effects of new
physics at an energy scale of $\Lambda$, much higher than the
electroweak scale, is the effective Lagrangian method. If the
underlying extended theory under consideration only becomes
important at a scale $\Lambda$, then it makes sense to expand the
Lagrangian in powers of $\Lambda^{-1}$:
\begin{eqnarray}
\mathcal{L} = \mathcal{L}_{SM} + \sum\frac{c_{i}}{\Lambda^{n_{i}-4}}O_{i}
\end{eqnarray}
where $\mathcal{L}_{SM}$ is the standard model Lagrangian,
$O_{i}$'s are the operators containing {\it only} the SM fields,
$n_{i}$ is the dimension of $O_{i}$ and $c_{i}$'s are
dimensionless parameters \cite{zhang},\cite{han},\cite{buchmuller}.

In the top quark sector, the lowest dimension operators that contribute
to FCNC with the $Z\bar{t}c$ vertex can be written as \cite{han2}:
\begin{eqnarray}\label{fcnc}
 \mathcal{L}_{eff} = -\frac{g}{2\cos\theta_{W}}\Bigl[\kappa_{L}Z^{\mu}\bar{t}
\gamma_{\mu}P_{L}c+\kappa_{R}Z^{\mu}\bar{t}
\gamma_{\mu}P_{R}c + h.c.\Bigr]
\end{eqnarray}
where $g$ is the coupling constant of $SU(2)_{L}$, $\theta_{W}$
is the Weinberg mixing angle and $\kappa_{L,R}$ are free
parameters determining the strength of these anomalous couplings.
Assuming CP invariance  $\kappa_{L,R}$ are real. In the above
relation, $P_{L,R}$ are the left-handed and right-handed
projection operators. The top FCNC anomalous interaction leads to
the following branching fraction for the $t\rightarrow Zc$ (in the
limit of zero mass of $b,c$ quarks):
\begin{eqnarray}
BR(t\rightarrow Zc)\equiv \frac{\Gamma(t\rightarrow
Zc)}{\Gamma(t\rightarrow Wb)} =
\frac{(\kappa_{L}^{2}+\kappa_{R}^{2})}{2}\frac{(m_{t}^{2}-m_{Z}^{2})^{2}(m_{t}^{2}+2m_{Z}^{2})}
{(m_{t}^{2}-m_{W}^{2})^{2}(m_{t}^{2}+2m_{W}^{2})}\simeq
0.5(\kappa_{L}^{2}+\kappa_{R}^{2})
\end{eqnarray}
Recent upper bound from CDF experiment for branching fraction
of $t\rightarrow Zq,(q=u,c)$, is $3.7\%$ (with $95\%$ C.L.)\cite{cdf}.
Therefore, one can conclude $\kappa_{L}^{2}+\kappa_{R}^{2} < 0.074$.

\section{Estimation of the Constraints}

We consider the effective interaction of the quarks
with on-shell photons to make prediction of the acceptable
range for the FCNC parameters ($\kappa_{L},\kappa_{R}$)
\cite{Hewett},\cite{edm1},\cite{edm2}:
\begin{eqnarray}\label{edmlag}
 \mathcal{L}_{eff} = -\frac{i}{2}d_{q}\bar{q}\gamma_{5}\sigma_{\mu\nu}qF^{\mu\nu}
\end{eqnarray}
where $F^{\mu\nu}$ is the electromagnetic field tensor and
$d_{q}$ is the top quark electric dipole moment (EDM) which is a
real number by hermiticity.
One should note that this is a CP violating term. Therefore, a
non-vanishing value for EDM of a fermion is of special interest
as it signifies the presence of CP violating interactions.

In the SM, top quark  can not have an EDM at least to three-loops.
The SM prediction for the top EDM is $10^{-31}-10^{-32}$ e-cm which is
too small to be observable. In contrast, in extensions of SM,
such as MSSM, this situation changes sharply and the top quark EDM
can arise at the one-loop. In beyond standard model theories the
typical top EDM is of order of $10^{-18}-10^{-20}$ e-cm which is
larger than the SM prediction by more than 10 orders of magnitude \cite{edm1},\cite{edm2}.

\begin{figure}
\centering
  \includegraphics[width=7cm,height=6cm]{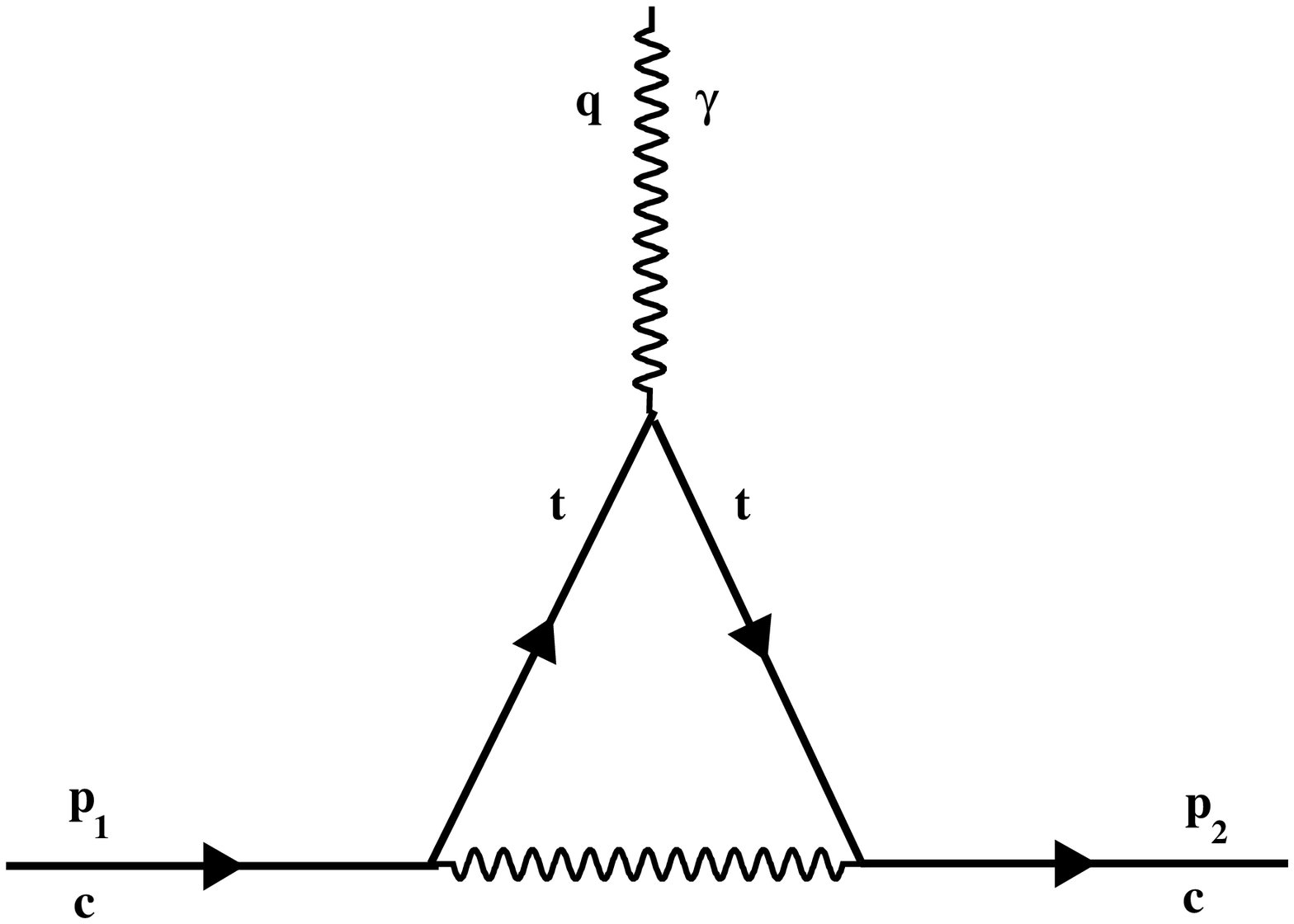}
  \includegraphics[width=7cm,height=6cm]{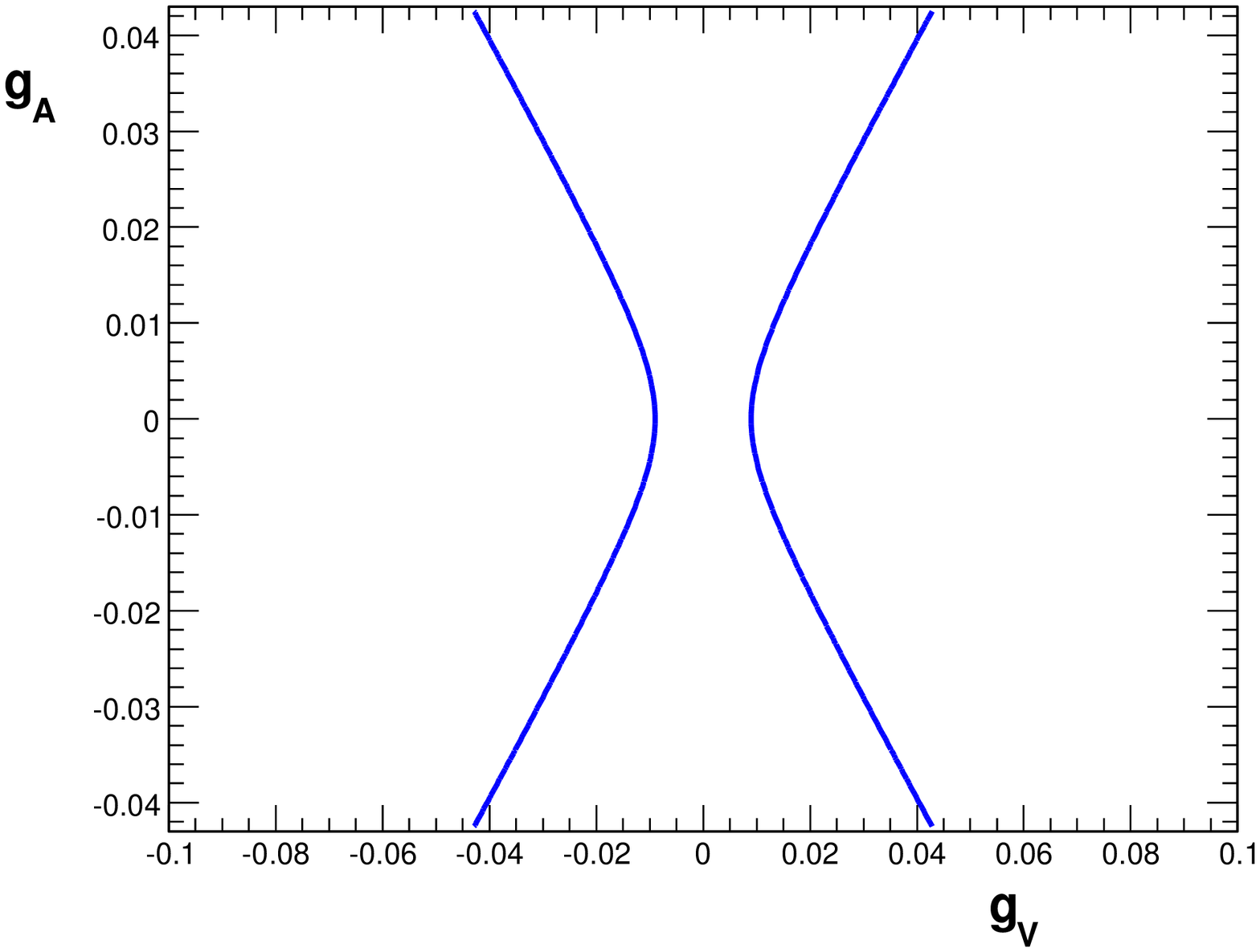}\\
  \caption{Left: The FCNC one-loop contribution to the vertex $\bar{c}c\gamma$, Right:
  The exclusion contour on top FCNC anomalous couplings.}\label{contour}
\end{figure}

The contribution of top quark FCNC introduced by Eq.\ref{fcnc} to
the on-shell $c\bar{c}\gamma$ coupling is given through the
diagram shown in the left side of Fig.\ref{contour}. Using the
effective interactions in Eq.\ref{fcnc}, Eq.\ref{edmlag} and after
some algebra the respective one-loop vertex can be written as:
\begin{eqnarray}
\Gamma_{\mu} &=& -\frac{g^{2}}{4\cos\theta^{2}_{W}}\times d_{t} \\ \nonumber &\times&
\int \frac{d^{4}k}{(2\pi)^{4}}\frac{\Bigl[...+\bigl(m_{t}(\kappa_{L}^{2}+\kappa_{R}^{2})(\PS_{2}-\PS_{1})-4\kappa_{L}\kappa_{R}(k.(p_{1}+p_{2})-p_{1}.p_{2}+m_{t}^{2}-m_{Z}^{2})\bigr)\gamma_{5}\sigma_{\mu\nu}q^{\nu} \Bigr]}{(k^{2}-m_{Z}^{2})
((p_{1}-k)^{2}-m_{t}^{2})((p_{2}-k)^{2}-m_{t}^{2})}
\end{eqnarray}

One should note that there are contributions to both magnetic and electric
dipole moments of the charm quark which we have kept only the terms contributing to
EDM. After evaluating the integral over $k$:
\begin{eqnarray}\label{edm}
d_{c} = \frac{\alpha}{4\pi\sin^{2}\theta_{W}\cos^{2}\theta_{W}}
\times d_{t}\times
[(\kappa_{L}^{2}+\kappa_{R}^{2})f(x_{t},x_{c})+\kappa_{L}\kappa_{R}g(x_{t})]
\end{eqnarray}
where
\begin{eqnarray}
f(x_{t},x_{c}) = \frac{\sqrt{x_{t}x_{c}}}{2}
\times\frac{1+3x_{t}(x_{t}-4/3)-2(2x_{t}-1)\log(x_{t})}{2(x_{t}-1)^{3}}~,
~g(x_{t}) = \frac{x_{t}-\log(x_{t})-1}{x_{t}-1}
\end{eqnarray}
where $x_{t} = \frac{m_{t}^{2}}{m_{Z}^{2}}$ and $x_{c} =
\frac{m_{c}^{2}}{m_{Z}^{2}}$. In obtaining the above relation, we
have ignored of the terms proportional to $(m_{c}^{2}/m_{Z}^{2})$,
which in fact is a negligible quantity.

In \cite{toscano2}, the authors have estimated the upper bound of
$7\times10^{-21}$ on the top and $1\times 10^{-27}$ on the charm
quark electric dipole moments using the experimental bound on
neutron electric dipole moment. The combination of these bounds
and Eq.\ref{edm} leads to the exclusion contour shown in the
right side of Fig.\ref{contour}. However, in this exclusion
contour another parameterizations for top FCNC anomalous coupling
has been used which are related to the parameters of
Eq.\ref{fcnc} by: $\kappa_{R,L} \equiv g_{V}\pm g_{A}$. If we
combine our result with  that obtained by CDF experiment, which
mentioned before, the bounds on  $\kappa_{R,L}$ are estimated as:
$\kappa_{L} < 3\times 10^{-3} ~,~ \kappa_{R} < 0.27$. These
values are compatible with the ones estimated in the other
studies \cite{han2}. The obtained upper bound on $\kappa_{L}$ in
\cite{han2} is $5\times 10^{-2}$. Hence, the allowed region for
$\kappa_{L}$ from the present work is one order of magnitude
smaller than the one obtained in \cite{han2}.

\section{Conclusion}

In this article, within the framework of the effective Lagrangian
approach, we performed a calculation of the radiative corrections
induced on the charm quark electric dipole moment by the
effective FCNC vertex $tcZ$. Using the present upper bounds on
the top and charm quark EDM's, the new limits on the top FCNC
anomalous couplings were estimated: $\kappa_{L} < 3\times
10^{-3}~,~ \kappa_{R} < 0.27$. These limits are regular and
comparable with the ones obtained in the past studies and the
estimated limit on $\kappa_{L}$ is slightly better.

{\large \bf Acknowledgments}\\
The authors would like to thank A. Khorramian for his supports.\\

\end{document}